\documentclass[12pt]{article}

\newcommand{\beq}{\begin{equation}}
\newcommand{\eeq}{\end{equation}}
\newcommand{\beqa}{\begin{eqnarray}}
\newcommand{\eeqa}{\end{eqnarray}}
\newcommand{\beqar}{\begin{eqnarray*}}
\newcommand{\eeqar}{\end{eqnarray*}}

\newcommand{\al}{\alpha}
\newcommand{\be}{\beta}

\def\spa          {\ \ \ }
\def\non          {\nonumber}
\def\ha           {\mbox{$\frac{1}{2}$}}

\def\spa          {\ \ \ }
\def\mand         {\spa\mbox{and}\spa}

\def\Tr           {\mbox{\rm Tr}\,}
\def\STr          {\mbox{\rm STr}\,}

\def\cd           {{\cdot}}
\def\ran          {\rangle}
\def\lan          {\langle}
\def\fsH	{H\!\!\!\!/\,}
\newcommand{\del}{\delta}

\newcommand{\eps}{\epsilon}
\newcommand{\ga}{\gamma}
\newcommand{\Ga}{\Gamma}

\newcommand{\inn}{\!\cdot\!}

\newcommand{\lam}{\lambda}

\newcommand{\z}{\zeta}

\newcommand{\labell}[1]{\label{#1}} 
\newcommand{\reef}[1]{(\ref{#1})}
\newcommand\prt{\partial}

\newcommand\bD{\bar{D}}

\def\fsC    {C\!\!\!\!/\,}

\begin{document}
\baselineskip 18pt%
\begin{titlepage}
\vspace*{1mm}%
\hfill%
\vspace*{13mm}%
\center{{\bf\Large On  D-brane -Anti D-brane Effective actions and their all order Bulk Singularity  Structures
}}
\begin{center}
{Ehsan Hatefi   \small $^{1,2,3}$}
\vspace*{0.04cm}
\vskip.1in
{ $^{1}$ Centre for Research in String Theory, School of Physics and Astronomy,
\\
Queen Mary University of London, Mile End Road, London E1 4NS, United Kingdom},
\vskip.06in
{ $^{2}$ Institute for Theoretical Physics, TU Wien
\\
Wiedner Hauptstrasse 8-10/136, A-1040 Vienna, Austria}
\vskip.06in
{ $^{3}$  ehsan.hatefi@tuwien.ac.at, ehsan.hatefi@cern.ch, e.hatefi@qmul.ac.uk}
\vspace*{0.01cm}
\end{center}
\begin{center}{\bf Abstract}\end{center}
\begin{quote}

All four  point functions of brane anti brane system including their correct all order $\alpha'$ corrections have been addressed. All five point functions of one closed string Ramond-Ramond (RR), two real  tachyons and either one gauge field or the scalar field in both symmetric and asymmetric pictures  have also been explored.  The entire analysis of  $<C^{-2} A^0  T^0 T ^{0}>$ is carried out.  Not only does it  fix the vertex operator of RR in asymmetric picture  and in higher point functions of string theory amplitudes but also it confirms the fact that  there is no  issue of picture dependence of the mixed closed RR, gauge fields,  tachyons and fermion fields in all symmetric or anti symmetric ones. We compute  $<C^{-2} \phi^0  T^0 T ^{0}>$ S-matrix in the presence of a transverse scalar field, two real tachyons  and that reveals two different kinds of bulk singularity structures, involving an infinite number of $u$- channel gauge field and $(u+s'+t')$ -channel scalar bulk poles. In order to produce all those bulk singularity structures, we define various couplings  at the level of the effective field theory that involve the mixing term of Chern-Simons 
 coupling (with C-potential field) and  a covariant derivative of the scalar field that comes from the pull-back of brane. Eventually  we explore their all order $\alpha'$ corrections in the presence of brane anti brane system where  various remarks will be also pointed out.

\end{quote}
\end{titlepage}

\section{Introduction}

By studying the unstable branes in detail , one may hope to have some sort of understanding about supersymmetry breaking as well as  getting more knowledge about at least some of the string theory 's properties in the 
backgrounds which are specially time dependent \cite{Gutperle:2002ai,Lambert:2003zr,Sen:2004nf}. 
\vskip.1in

Having dealt with unstable branes , one could also  talk about Sakai- Sugimoto model \cite{Sakai:2004cn},  the  low energy hadron physics in some of the holographic QCD models  as well as  studying properly the so called  spontaneous symmetry breaking  in  some of the holographic QCD models \cite{Casero:2007ae}. Indeed  we try to define 
flavour branes by embedding various parallel branes and consider all anti branes inside a  background which must be  dual to some theory such as colour confined theorem. In fact one might consider brane anti brane like system as a probe if and only if $N_f<< N_c$.
Strictly  speaking,  one reveals the fact that  the presence of tachyons in the spectrum  is the only source of instability in  all these formalisms, therefore it absolutely  makes sense to actually work out within details with some field theories that do include these unstable modes. 

\vskip.1in

It was A.Sen who realized an effective actions \cite{Sen:1999md,Bergshoeff:2000dq,Kluson:2000iy} that involves  tachyons, can describe various issues  such as the decays  \cite{Sen:2002in} of 
the odd parity  branes  or the so called non-BPS branes in superstring theory.  It is also shown in  \cite{Sen:2004nf} that the Effective Field Theory (EFT)  for all non-BPS branes includes just  massless states as well as  tachyons as we have dealt with 
non-BPS  effective actions in \cite{Garousi:2008ge} where various remarks on  tachyon condensation for brane anti brane system  have also been presented in \cite{Sen:1998sm}.

\vskip.1in

Sen has also  clarified  for brane anti brane system   in \cite{Sen:1998ii}  that once  the separation of branes becomes smaller than string length scale then two real tachyon 
modes show up in the spectrum of brane anti brane , hence it makes sense to substitute them in an effective action of brane anti brane  system as they are going to play for us the most fundamental role in ascribing the dynamics of the configuration.

Recently in \cite{Michel:2014lva} various remarks about the dynamic of brane anti brane system  have been made, namely it is discussed that if one employs the brane actions  in the context
of EFT  , then one is able to even properly deal with loop divergences. 
\vskip.1in

Let us address various motivations or applications for brane anti brane systems .    

\vskip.1in

Brane production  \cite{Bergman:1998xv},  dealing with stable nonBPS D-branes in all type I,II string theory  as well as  having inflation in string theory  in the language of brane anti brane or KKLT   are suggested  \cite{Dvali:1998pa}.

Note that within just  S-Matrix computations one assured that not only various new couplings can be verified but also all the coefficients of the string theory couplings are exact and appear without any ambiguity.\footnote{ Several remarks on higher point functions of the string amplitudes \cite{Bjerrum-Bohr:2014qwa} as well as their corrections in  \cite{Hatefi:2012zh,Hatefi:2012ve} are made , we  indeed derived 
all order $\alpha'$  corrections to BPS branes , where one might be interested in looking at the eventual conjecture \cite{Hatefi:2012rx} on $\alpha'$ corrections that has worked out surprisingly for both non-BPS and supersymmetric cases.}Just as an extra comment,  one could consider the thermodynamical aspects of brane anti brane action, because it is known that  at finite temperature this D-brane anti D-brane  could be stabilised and is kind of related to black holes where  to our knowledge some of
the applications to   AdS/CFT or  to M-theory \cite{Hatefi:2012bp} were appeared in the literature.

\vskip.1in

It is worth pointing out the fact that brane anti brane has been playing a very strong role for the stability of KKLT as well as in large volume scenario \cite{Polchinski:2015bea}.  
\vskip.1in

In   \cite{Polchinski:1995mt}   Polchinski  explained  the deep and close relationship between the D-branes and   the closed string RR field  where we just demonstrate the very main paper on the content of bound states of the branes
\cite{Witten:1995im}. One reason in favour of scattering amplitude calculations is indeed its strong potential in getting exact and correct $\alpha'$ corrections.  Although there is no duality transformation for non-BPS branes, it is worth to stress the following remark. In fact it was shown that  even for  various BPS S-matrices  all order $\alpha'$ corrections of DBI action with exact coefficients can not be fixed by just applying T-duality transformation. Indeed it is clarified that just  with direct S-matrix calculations, one can hope to  precisely gain all order $\alpha'$ corrections of DBI action with exact coefficients for instance one might look at   \cite{Hatefi:2012zh} .

\vskip .1in

To be able to read off all the ingredients of either Wess-Zumino (WZ) or DBI  effective actions , we suggest \cite{Hatefi:2010ik,Myers:1999ps}.


 \vskip.2in

The  paper is based on the following contents. In section two, we do evaluate the amplitude of two real tachyons of brane anti brane and a C - field ( the so called RR potential term  in asymmetric picture of the closed string RR),
where this S-matrix in symmetric picture,  at leading order  was computed in \cite{Kennedy:1999nn}.  We expand it out  and explore not only its low energy limit by re generating a gauge field pole  but also find out its correct 
and exact all infinite $\alpha'$ higher derivative corrections  of two tachyons and a C-field.  Basically we generate  all order $\alpha'$  contact terms of $C_{p-1}\wedge dT\wedge dT^*$.
In section three, we  compute $<C^{-2} A^{0} T^0 T^0>$ amplitude in detail  and talk about its unique expansion, we then start comparing it with its symmetric result. The outcome for both symmetric and asymmetric picture is the same and 
for the first time we understand the fact that even in higher point functions of string theory there is no picture dependence for the mixed RR and world volume strings, such as gauge fields, tachyons  and fermions but not scalar fields.

\vskip.1in

This evidently confirms that the vertex operator of RR in asymmetric picture is exact and complete and no extra potential terms needed to be added to RR potential vertex operator  in asymmetric picture.  It also confirms that for mixed world volume S-matrices of closed string RR- gauge field and tachyon , even in five and higher point functions there is no picture dependence at all. However,  the story gets complicated for the higher point functions of the mixed closed string RR and scalar fields as we point out in detail in  section four. In section five we point out $<C^{-2} \phi^{0} T^0 T^0>$ as well as its all two different symmetric results.
 Two different kinds of  singularity structures  that carry the scalar product of momentum of  C-field in the bulk  and scalar polarization $(p^i\xi_{1i})$ will be obtained.  These terms can not be derived by momentum conservation along the world volume of brane and we called them bulk singularities.  In order to produce those bulk singularities  in an effective field theory we propose  a new sort of coupling as follows 
 
 \beqa
 \frac{1}{(p-1)!}\mu_p (2\pi\alpha')^2\int_{\sum_{(p+1)}}\Tr\bigg( C^{i}_{a_0...a_{p-3}} F_{a_{p-2}a_{p-1}} D_{a_{p}}\phi^i\bigg)\labell{newrr22oi}
 \eeqa

where in above the scalar field comes from pull-back of brane and the Chern-Simons coupling  has also been taken into account. Setting \reef{newrr22oi}  and discovering 
its all order $\alpha'$ corrections, we are able to precisely produce an infinite number of $u$ channel bulk singularities of this  S-matrix. Eventually we introduce a new coupling as follows
\beqa
(2\pi\alpha')\mu_p\frac{1}{(p+1)!} \int_{\Sigma_{p+1}}\eps^{a_0\cdots
a_{p}}
 C^{i }_{a_0\cdots a_{p-1}} D_{a_{p}} \phi^i \label{esi897}\eeqa

where in the above  coupling,  the scalar field has been taken from pull back of brane. We then derive its vertex operator in an EFT and by making use of all order $\alpha'$ corrections of two tachyon two scalar couplings in the world volume of brane anti brane  \cite{Hatefi:2012cp} we exactly reconstruct all infinite $(u+s'+t')$- channel bulk poles of this string amplitude in the background of brane anti brane in an EFT as well. 
It is worth pointing out that, within direct scattering computations in \cite{Hatefi:2012cp}, we have already found several new couplings, such as $D\phi^{1i}.D\phi_{2i}$ in brane anti brane system that shall be used in this paper as well.
Let us turn to actual details of the paper.

\section{The $C^{-2}-T^{0}-T^0$ S-matrix}

In this section first we would like to  just mention very briefly  the effective actions of brane anti brane system and then start analyzing the higher point functions in that background.

All the effective actions of a $D_p\bD_p$-brane in both Type IIA(B) theory can be achieved by  embedding tachyons in both DBI  and  Wess-Zumino (WZ) actions. One can consider just two  non-BPS D-branes in Type IIB(A) and project them out, 
however,  for the purpose of this paper we limit ourselves to two tachyons,  either a gauge field or an scalar field and an RR where RR part comes from Chern-Simons action and the other fields do appear in DBI as follows \cite{Garousi:2007fk}
\beqa
S_{DBI}&=&-\int
d^{p+1}\sigma \Tr\left(V({\cal T})
\sqrt{-\det(\eta_{ab}
+2\pi\alpha'F_{ab}+2\pi\alpha'D_a{\cal T}D_b{\cal T})} \right)\,\,,\labell{nonab} \eeqa 
Note that the trace in \reef{nonab} must be symmetric for all the matrices  such as 
$F_{ab},D_a{\cal T}$, 
${\cal T}$ inside the potential that have shown up in the action. \footnote{ The definitions for all the matrices  are
\beqa
F_{ab}=\pmatrix{F^{(1)}_{ab}&0\cr 
0&F^{(2)}_{ab}},\,\,
D_{a}{\cal T}=\pmatrix{0&D_aT\cr 
(D_aT)^*&0},\,\, {\cal T}=\pmatrix{0&T\cr 
T^*&0}\,\labell{M12} \eeqa 
 with $F^{(i)}_{ab}=\prt_{a}A^{(i)}_{b}-\prt_{b}A^{(i)}_{a}$ and $D_{a}T=\prt_{a}T-i(A^{(1)}_a-A^{(2)}_a)T$.}
 
 \vskip.1in
 
Suppose  we use the ordinary trace, then \reef{nonab} gets replaced by A. Sen 's action \cite{Sen:2003tm}, if we try to make  the kinetic term symmetrized and evaluate the trace, however, there is an important fact  which one must be aware of as follows.
 \vskip.1in
 
 Indeed  it is shown by direct analysis of scattering amplitude in \cite{Hatefi:2012cp,Garousi:2007fk} that Sen's effective action is  not  consistent  with string theory amplitudes. Tachyon 's consistent potential  in the context of scattering amplitude is 
 \beqa
V(|T|)&=&1+\pi\alpha'm^2|T|^2+
\frac{1}{2}(\pi\alpha'm^2|T|^2)^2+\cdots
\non\eeqa 

with $m^2=-1/(2\alpha')$ is tachyons's mass square and  $T_{p}$ is the
tension of a p-brane. Notice that the expansion has given us a very consistent result for the 
potential of $V(|T|)=e^{\pi\alpha'm^2|T|^2}$ that has been imposed by boundary string field theory (BSFT) as well \cite{Kutasov:2000aq}.
 
  \vskip.1in
  
 Note that in an important paper \cite{Tseytlin:2000mt} a sigma model approach to string theory effective actions with tachyons,
has also been released.  It is discussed in detail in   \cite{Garousi:2007fk} that only the above effective action which is based on the direct string theory S-matrix calculations can produce all the infinite poles and contact interactions where the mixing terms such as $F^{(1)}\cdot{F^{(2)}}$ and $D\phi^{(1)}\cdot{D\phi^{(2)}}$ play the crucial roles in getting consistent results between field theory and string amplitudes so that the Lagrangian that contributes to an S-matrix of  a gauge field and two
tachyons  (in the presence of RR)  does involve the following interactions \cite{Garousi:2007fk}: 
\beqa {\cal
L}_{DBI}&\!\!\!=\!\!\!&-T_p(2\pi\alpha')\left(m^2|T|^2+DT\cdot(DT)^{*}-\frac{\pi\alpha'}{2}
\left(F^{(1)}\cdot{F^{(1)}}+
F^{(2)}\cdot{F^{(2)}}\right)\right)+T_p(\pi\alpha')^3\nonumber\\
&&\times\left(\frac{2}{3}DT\cdot(DT)^{*}\left(F^{(1)}\cdot{F^{(1)}}+F^{(1)}\cdot{F^{(2)}}+F^{(2)}\cdot{F^{(2)}}\right)\right.\labell{exp1}\\
&&\left.+\frac{2m^2}{3}|\tau|^2\left(F^{(1)}\cdot{F^{(1)}}+F^{(1)}\cdot{F^{(2)}}+F^{(2)}\cdot{F^{(2)}}\right)\right.\nonumber\\
&&-\left.\frac{4}{3}\left((D^{\mu}T)^*D_{\beta}T+D^{\mu}T(D_{\beta}T)^*\right)\left({F^{(1)}}^{\mu\alpha}F^{(1)}_{\alpha\beta}+{F^{(1)}}^{\mu\alpha}F^{(2)}_{\alpha\beta}+{F^{(2)}}^{\mu\alpha}F^{(2)}_{\alpha\beta}\right)\right)
\nonumber
\eeqa

On the other hand the  WZ action of brane anti brane system that defines for us the allowed couplings of RR field to gauge field is derived in \cite{Douglas:1995bn}\footnote{
Note that $C$ is a  sum on  RR potentials $C=\sum_n(-i)^{\frac{p-m+1}{2}}C_m$. 
 }
\beqa
S = \mu_p\int_{\Sigma_{(p+1)}} C \wedge  \left(e^{i2\pi\alpha'F^{(1)}}-e^{i2\pi\alpha'F^{(2)}}\right)\ ,
\labell{eqn.wz}
\eeqa

In  \cite{Kraus:2000nj} it is shown how to impose the tachyons inside the actions also one may use  the so called super-connection of non-commutative geometry ~\cite{quil,berl,Roepstorff:1998vh} as follows 
\beqa
S_{WZ}&=&\mu_p \int_{\Sigma_{(p+1)}} C \wedge \STr e^{i2\pi\alpha'\cal F}\labell{WZ}\eeqa \footnote{ 
with some  definitions for the curvature 
\beqa {\cal F}&=&d{\cal A}-i{\cal A}\wedge\cal A\nonumber \eeqa
and the super-connection as \begin{displaymath}
i{\cal A} = \left(
\begin{array}{cc}
  iA^{(1)} & \beta T^* \\ \beta T &   iA^{(2)} 
\end{array}
\right) \ ,
\non\end{displaymath}}
where it is shown in  \cite{Garousi:2007fk} how to get to  the curvature as below 
\begin{displaymath}
i{\cal F} = \left(
\begin{array}{cc}
iF^{(1)} -\beta^2 |T|^2 & \beta (DT)^* \\
\beta DT & iF^{(2)} -\beta^2|T|^2 
\end{array}
\right) \ ,
\non\end{displaymath}
with $F^{(i)}=\frac{1}{2}F^{(i)}_{ab}dx^{a}\wedge dx^{b}$ and $DT=[\partial_a T-i(A^{(1)}_{a}-A^{(2)}_{a})T]dx^{a}$. Essentially one can extract the terms inside the WZ action \reef{WZ} to reconstruct the following terms that are needed later on,
\beqa
C\wedge \STr i{\cal F}&\!\!\!\!=\!\!\!&C_{p-1}\wedge(F^{(1)}-F^{(2)})\labell{exp2}\\
C\wedge \STr i{\cal F}\wedge i{\cal F}&\!\!\!\!=\!\!\!\!&C_{p-3}\wedge \left\{F^{(1)}\wedge F^{(1)}-
F^{(2)}\wedge F^{(2)}\right\}\nonumber\\
&& +C_{p-1}\wedge\left\{-2\beta^2|T|^2(F^{(1)}-F^{(2)})+2i\beta^2 DT\wedge(DT)^*\right\}\nonumber\\
\eeqa

Let us turn to the calculations.

\vskip.1in

First of all note that the space-time two point function or world sheet three point function of one closed string RR and a real tachyon of non-BPS branes in both asymmetric and symmetric pictures in  type IIA and IIB has already been calculated in \cite{Hatefi:2015gwa} as follows:
\beqa
{\cal A}^{C^{-1}T^{-1}} & \sim & -2i \Tr(P_{-}\fsH_{(n)}M_p)
\nonumber\\
{\cal A}^{C^{-2}T^{0}} & \sim & 2^{1/2} (2ik_{1a})\Tr(P_{-}\fsC_{(n-1)}M_p\gamma^a)
 \nonumber\eeqa
Applying the momentum conservation on the world volume , $(k_1+p)^a=0 $, making use of  $p^a\fsC_{n-1}=\fsH_n$ 
we are able to produce the same result in both pictures. If we would actually use 
 \beqa
 2i\pi\alpha'\beta'\mu'_p\int C_p\wedge DT\nonumber\eeqa
  coupling in field theory, then  we would be able to precisely produce the whole S-matrix of string amplitude (in this case is just pure contact interaction) in effective theory as well. Let us turn to world sheet  four point function.

  \vskip.1in
  
Here we do want to make use of CFT methods \cite{Friedan:1985ge} to actually gain the four point function of two real tachyons and one asymmetric closed string RR which makes sense in the presence of D-brane anti D-brane of type II systems. The  structure  of $C^{-2}$ has been first pointed out  in \cite{Bianchi:1991eu} and then later  was accommodated by \cite{Liu:2001qa} so that

\beqa
{\cal A}^{C^{-2}T^0 T^0} & \sim & \int dx dy d^2z
 \lan
V_{T}^{(0)}(y)V_T^{(0)}(x)
V_{RR}^{(-2)}(z,\bar{z})\ran,\labell{cor1}\eeqa
where  all the vertices including their Chan-Paton factors in brane anti brane system can be achieved  as
 \beqa
V_{T}^{(0)}(y) &=&\alpha'ik_1\cd\psi(y) e^{\alpha'ik_1.X(y)}\lam\otimes\sigma_1 \nonumber\\
V_{RR}^{(-2)}(z,\bar{z})&=&(P_{-}\fsC_{(n-1)}M_p)^{\al\be}e^{-3\phi(z)/2} S_{\al}(z)e^{i\frac{\alpha'}{2}p\cd X(z)}
e^{-\phi(\bar{z})/2} S_{\be}(\bar{z}) e^{i\frac{\alpha'}{2}p\cd D \cd X(\bar{z})}\otimes I\label{vertices}\eeqa

where at disk level amplitude both tachyons must be replaced on the boundary while RR is located at the middle of the defined disk.
\vskip .2in

On-shell conditions as well as the other definitions are   

\beqa
   p^2=0, \quad  k_{1}^2=k_{2}^2=1/4 
\nonumber\\
P_{-} =\ha (1-\ga^{11}), \quad
\fsH_{(n)} = \frac{a
_n}{n!}H_{\mu_{1}\ldots\mu_{n}}\ga^{\mu_{1}}\ldots
\ga^{\mu_{n}},
\nonumber\\
(P_{-}\fsH_{(n)})^{\al\be} =
C^{\al\del}(P_{-}\fsH_{(n)})_{\del}{}^{\be}.
\eeqa

where  $n=2,4$,$a_n=i$  ($n=1,3,5$,$a_n=1$) holds for   type IIA  (type IIB).

Let us just deal with the holomorphic counterparts of  world-sheet fields, to do so we apply the so called doubling trick, which means that the following change of variables has been taken into account
\begin{displaymath}
\tilde{X}^{\mu}(\bar{z}) \rightarrow D^{\mu}_{\nu}X^{\nu}(\bar{z}) \ ,
\spa
\tilde{\psi}^{\mu}(\bar{z}) \rightarrow
D^{\mu}_{\nu}\psi^{\nu}(\bar{z}) \ ,
\spa
\tilde{\phi}(\bar{z}) \rightarrow \phi(\bar{z})\,, \mand
\tilde{S}_{\al}(\bar{z}) \rightarrow M_{\al}{}^{\be}{S}_{\be}(\bar{z})
 \ ,
\non\end{displaymath}

with the aforementioned matrices as below
\begin{displaymath}
D = \left( \begin{array}{cc}
-1_{9-p} & 0 \\
0 & 1_{p+1}
\end{array}
\right) \ ,\,\, \mand
M_p = \left\{\begin{array}{cc}\frac{\pm i}{(p+1)!}\ga^{i_{1}}\ga^{i_{2}}\ldots \ga^{i_{p+1}}
\eps_{i_{1}\ldots i_{p+1}}\,\,\,\,{\rm for\, p \,even}\\ \frac{\pm 1}{(p+1)!}\ga^{i_{1}}\ga^{i_{2}}\ldots \ga^{i_{p+1}}\ga_{11}
\eps_{i_{1}\ldots i_{p+1}} \,\,\,\,{\rm for\, p \,odd}\end{array}\right.
\non\end{displaymath}
\vskip .2in

Having set them , we are admitted from now on to indeed just work out with  holomorphic parts of the propagators for all world sheet fields of  $X^{\mu},\psi^\mu, \phi$ as below 
\begin{eqnarray}
\lan X^{\mu}(z)X^{\nu}(w)\ran & = & -\frac{\alpha'}{2}\eta^{\mu\nu}\log(z-w) \ , \non \\
\lan \psi^{\mu}(z)\psi^{\nu}(w) \ran & = & -\frac{\alpha'}{2}\eta^{\mu\nu}(z-w)^{-1} \ ,\non \\
\lan\phi(z)\phi(w)\ran & = & -\log(z-w) \ .
\labell{prop2}\end{eqnarray}

The CP factor for the above S-matrix is now  $2\Tr(\lam_1\lam_2)$. 
Having replaced the vertices inside the S-matrix, the amplitude can be written down as 
\beqa
&&-2\Tr(\lam_1\lam_2)\alpha'^2 k_{1a} k_{2b}\int dx_1 dx_2 dx_4 dx_5    (x_{45})^{-3/4} I_1
(P_{-}\fsC_{(n-1)}M_p)^{\alpha\beta} \nonumber\\&&\times
<:S_{\al}(x_4): S_{\be}(x_5):\psi^{a}(x_1):\psi^{b}(x_2):>,
 \nonumber\eeqa
where $x_4=z=x+iy,x_5=\bar z=x-iy$ and $x_{ij}=x_i-x_j$

\beqa
I_1&=&|x_{12}|^{\alpha'^2k_1.k_2}|x_{14}x_{15}|^{\frac{\alpha'^2}{2}k_1.p} |x_{24}x_{25}|^{ \frac{\alpha'^2}{2}  k_2.p}|x_{45}|^{\frac{\alpha'^2}{4}p.D.p},\nonumber\\
\eeqa

 One has to use  the modified  Wick-like rule \cite{Hatefi:2015jpa}  to be able to  derive the following correlation function
  \beqa
  <:S_{\al}(x_4): S_{\be}(x_5):\psi^{a}(x_1):\psi^{b}(x_2):>&=& \bigg((\Gamma^{ba} C^{-1})_{\alpha\beta}-2\eta^{ab} \frac{Re[x_{14}x_{25}]}{x_{12}x_{45}}\bigg)\nonumber\\&&\times2^{-1}(x_{14}x_{15}x_{24}x_{25})^{-1/2}(x_{45})^{-1/4}  \nonumber\eeqa

 If we insert the above correlator into the S-matrix element, one can easily show that the amplitude is  $SL(2,R)$ invariant. We do gauge fixing  by setting   $(x_1,x_2,z,\bar z)=(x,-x,i,-i)$ and taking $u = -\frac{\alpha'}{2}(k_1+k_2)^2$, the final outcome for the amplitude is \footnote{ $\alpha'=2$ is set. }
\beqa
{\cal A}^{C^{-2}T^{0}T^{0}} & \sim & 16i \Tr(\lam_1\lam_2) k_{1a}k_{2b}\int_{-\infty}^{\infty} dx (2x)^{-2u-1}
\bigg((1+x^{2})\bigg)^{2 u} \nonumber\\&&\times\bigg[\Tr
(P_{-}\fsC_{(n-1)}M_p\Gamma^{ba})-2\eta^{ab}\frac{1-x^2}{4xi}\Tr(P_{-}\fsC_{(n-1)}M_p)\bigg],\nonumber\eeqa

obviously the second term in above does not have any contribution to the entire S-matrix and the ultimate result reads off as

 \beqa
{\cal A}^{C^{-2}T^{0}T^{0}} &=&  \frac{i\mu_p}{4}   \Tr(\lam_1\lam_2)2\pi  k_{1a}k_{2b}\frac{\Gamma(-2u)}{\Gamma(1/2-u)^2}\Tr(P_{-}\fsC_{(n-1)}M_p\Gamma^{ba}) 
 \label{yy12}\eeqa

 meanwhile this amplitude in symmetric picture has been done in \cite{Kennedy:1999nn} as follows
 \beqa
{\cal A}^{C^{-1}T^{-1}T^{0}} & \sim&   \Tr(\lam_1\lam_2)2\pi \frac{\Gamma(-2u)}{\Gamma(1/2-u)^2}\Tr(P_{-}\fsH_{(n)}M_p\gamma^a)k_{2a}
 \label{yy1}\eeqa

\vskip.1in

If we use
the  momentum conservation  as $-k_1^{a} - p^{a} =k_2^{a}$ and more crucially apply  the following Bianchi identity $p_a \eps^{a_{0}...a_{p-1}a}=0$, then we get to know that  the S-matrix is antisymmetric under interchanging tachyons.   Note that to make sense of the amplitude in asymmetric picture , $p_a \eps^{a_{0}...a_{p-2}ba}$ should  be non-vanishing.


The amplitude is normalized   by  $ (i\mu_p/4)$ where  $ \beta $ is WZ constant and   $  \mu_p $ becomes brane's RR charge.  The trace can be extracted as   
\beqa
\Tr\bigg(\fsC_{(n-1)}M_p
(\Gamma^{ba})\bigg)\delta_{p,n}&=&\pm\frac{32}{(p-1)!}\eps^{a_{0}\cdots a_{p-2}ba}C_{a_{0}\cdots a_{p-2}}  \delta_{p,n}\nonumber\eeqa

  The trace including $\gamma^{11}$ confirms to us the fact that all the above results keep working for the following  as well
  
\beqa
  p>3 , H_n=*H_{10-n} , n\geq 5.
  \nonumber\eeqa

 It is discussed in  \cite{Hatefi:2012wj} that  by sending either $k_i.k_j\rightarrow 0$ or $(k_i+k_j)^2\rightarrow 0$, one finds out all massless or even tachyon singularities of BPS or non-BPS S-matrices. 
 We know that there is a non-zero coupling between   $C_{p-1}\wedge F$, two tachyons and a gauge field coupling, hence the correct momentum expansion for this amplitude 
 is \beqa
u=-p^ap_a\rightarrow 0.
\eeqa
 which is consistent with the fact that $p^ap_a$ has to be sent to zero for brane -anti brane systems \cite{Hatefi:2012cp}, while  for non- BPS branes the constraint  gets replaced by  $-p^ap_a\rightarrow \frac{-1}{4}$.
 
\vskip .2in

   Let us consider the expansion of the pre-factor in string amplitude as 
   \beqa
2\pi\frac{\Ga(-2u)}{\Ga(1/2-u)^2}
 &=& -\frac{1}{u}+\sum_{m=-1}^{\infty}c_m(u )^{m+1}
\ .\labell{taylor61}
\eeqa
with the following coefficients
\beqa
 c_{-1}&=&4ln(2), c_0=(\frac{\pi^2}{6}-8ln(2)^2),\nonumber\\
c_1&=&\frac{2}{3}(3\z(3)-\pi^2ln(2)+16ln(2)^3)
\nonumber\eeqa
   
   now we just want to produce the only massless gauge pole of this four point function by the following  sub-amplitude
\beqa
{\cal A}&=&V_a(C_{p-1},A^{(1)})G_{ab}(A)V_b(,A^{(1)},T_1,T_2)+V_a(C_{p-1},A^{(2)})G_{ab}(A)V_b(,A^{(2)},T_1,T_2)\labell{amp2}\eeqa
 First we need to take into account the following Chern-Simons coupling on the world volume of brane anti-brane system
 \beqa
  i\mu_p (2\pi\alpha')\int_{\Sigma_{p+1}} \epsilon^{a_0...a_{p}}\bigg(\Tr(C_{a_0...a_{p-2}}d_{a_{p-1}}(A_{1a_{p}}-A_{2a_{p}}))\bigg),
    \label{jj12}\eeqa

 Note that the off-shell gauge field  has to be $A^{(1)}$ and $A^{(2)}$ to actually get the consistent result  so all the field theory vertices and propagators are 
\beqa
G_{ab}(A) &=&\frac{i\delta_{ab}}{(2\pi\alpha')^2 T_p
\left(k^2\right)}\nonumber\\
V_b(A^{(1)},T_1,T_2)&=&iT_p(2\pi\alpha')(k_1-k_2)_{b}\nonumber\\
V_b(A^{(2)},T_1,T_2)&=&-iT_p(2\pi\alpha')(k_1-k_2)_{b}\nonumber\\
V_a(C_{p-1},A^{(1)})&=&i\mu_p(2\pi\alpha')\frac{1}{(p-1)!}\epsilon_{a_0\cdots a_{p-1}a}C^{a_0\cdots a_{p-2}} k^{a_{p-1}}\nonumber\\
V_a(C_{p-1},A^{(2)})&=&-i\mu_p(2\pi\alpha')\frac{1}{(p-1)!}\epsilon_{a_0\cdots a_{p-1}a}C^{a_0\cdots a_{p-2}}k^{a_{p-1}}\eeqa
Considering the momentum conservation  $k^a=(k_1+k_2)^{a}=-p^{a}$ and substituting them in  \reef{amp2}, we actually obtain precisely the same string elements in  the field theory amplitude as follows
\beqa
{\cal A}&=&4i\mu_p\frac{1}{(p-1)!u}\epsilon^{a_0\cdots a_{p-2}ab}C_{a_0\cdots a_{p-2}}k_{1a} k_{2b}\labell{amp444}\eeqa
Lets generate all the infinite contact interactions.

If we start expanding out all the Gamma functions inside the amplitude, we then derive all infinite $\alpha'$ higher derivative corrections to two real tachyons of brane anti brane system and a $C_{p-1}$ potential field. Consider the following coupling
  \beqa
    2i\beta^2\mu_p (2\pi\alpha')^2\int_{\Sigma_{p+1}} \bigg(\Tr(C_{p-1}\wedge DT\wedge DT^*)\bigg),
    \label{jj2}\eeqa

 now all order  $\alpha'$ higher derivative corrections to \reef{jj2}  are derived by taking into account $u\rightarrow 0$ in the amplitude.
 
 \vskip.1in

 Replacing the expansion to S-matrix and making the comparisons  with EFT coupling \reef{jj2} , we believe that  $c_{-1}$ term is regenerated by \reef{jj2}  so that the first contact term of string amplitude is

 \beqa
i\mu_p 16 ln2\frac{1}{(p-1)!}\epsilon^{a_0\cdots a_{p-2}ba}C_{a_0\cdots a_{p-2}} k_{1a} k_{2b}\labell{amp42p}\eeqa
if we compare \reef{amp42p} with \reef{jj2} then we can realise that, this  fixes the normalisation of WZ as 
$\beta=\frac{1}{\pi} \sqrt{\frac{2\ln(2)}{\alpha'}}$ and
 the second contact interaction becomes 
 \beqa
i\mu_p 4u \left(\frac{\pi^2}{6}-8\ln(2)^2\right)\frac{1}{(p-1)!}\epsilon^{a_0\cdots a_{p-2}ba}C_{a_0\cdots a_{p-2}} k_{1a} k_{2b}\labell{amp424}\eeqa
 where this can be produced by the following $\alpha'$ higher derivative corrections as 
 
\beqa
i(\alpha')^2\mu_p\left(\frac{\pi^2}{6}-8\ln(2)^2\right) \epsilon^{a_0\cdots a_{p}}C_{a_0\cdots a_{p-2}} (D ^b D_b) D_{a_{p-1}}T  D_{a_{p}}T^*\labell{hderv}
\eeqa
  
 One can keep finding out all  the other higher derivative corrections of \reef{jj2}  so that  all order $\alpha'$ corrections of one RR potential field and two real tachyons of brane anti brane system within compact formula  can be explored  as  follows
\beqa
i\mu_p (2\pi\alpha')^2  C_{(p-1)}\wedge \Tr\left(\sum_{m=-1}^{\infty}c_m(\alpha' (D ^b D_b))^{m+1}  DT \wedge DT^*\right) \labell{highaaw3}\eeqa
notice that  due to the constraint of producing the correct momentum expansion of brane anti brane system which is $u =-p^ap_a \rightarrow 0$ , the following corrections 
\beqa
i\mu_p (2\pi\alpha')^2  C_{(p-1)}\wedge \Tr\left(\sum_{m=-1}^{\infty}c_m(\alpha')^{m+1}  D_{a_1}\cdots D_{a_{m+1}}DT \wedge D^{a_1}...D^{a_{m+1}}DT^*\right) \labell{highaaw67}\eeqa

cannot be entirely worked out and indeed the only exact and correct all order $\alpha'$ corrections of this system is indeed \reef{highaaw3}, it is worth highlighting the essential point  as follows.  We were looking for two tachyons and a RR potential C field in the world volume of brane anti brane systems, therefore in above corrections all commutator terms must be overlooked.

\section{  All order  $<C^{-2} A^0 T^0 T^{0}>$ S-Matrix }

In this section we would like to actually carry out the entire S-matrix of a potential C-field, one gauge field and two real tachyons in the world volume of brane anti brane system to compare it with the same S-matrix in symmetric picture and eventually see whether or not one needs to add extra potential terms (just in the presence of world volume fields such as gauge field and tachyons ) to the vertex operator of RR in asymmetric picture. 
All the vertices in the presence of brane anti brane  system within their CP factor can be constructed as 

\beqa
V_{T}^{(0)}(x) &=&  \alpha' ik\cd\psi(x) e^{\alpha' ik\cd X(x)}\lam\otimes\sigma_1,
\nonumber\\
V_{T}^{(-1)}(x) &=&e^{-\phi(x)} e^{\alpha' ik\cd X(x)}\lam\otimes\sigma_2\nonumber\\
V_A^{(-1)}(x)&=&e^{-\phi(x)}\xi_a\psi^a(x)e^{ \alpha'iq\inn X(x)}\lam\otimes \sigma_3 \nonumber\\
V_{A}^{(0)}(x) &=& \xi_{1a}(\partial^a X(x)+i\alpha'k.\psi\psi^a(x))e^{\alpha'ik.X(x)}\otimes I \label{d4Vs}\\
V_{C}^{(-\frac{3}{2},-\frac{1}{2})}(z,\bar{z})&=&(P_{-}\fsC_{(n-1)}M_p)^{\alpha\beta}e^{-3\phi(z)/2}
S_{\al}(z)e^{i\frac{\alpha'}{2}p\cd X(z)}e^{-\phi(\bar{z})/2} S_{\be}(\bar{z})
e^{i\frac{\alpha'}{2}p\cd D \cd X(\bar{z})}\lam\otimes I \nonumber
\eeqa

This S-matrix is given by  
\beqa
{\cal A}^{C^{-2} A^0 T^0 T^0} & \sim & \int dx_{1}dx_{2}dx_{3}dzd\bar{z}\,
  \lan V_{A}^{(0)}{(x_{1})}
V_{T}^{(0)}{(x_{2})}V_{T}^{(0)}{(x_{3})}
V_{RR}^{(-\frac{3}{2},-\frac{1}{2})}(z,\bar{z})\ran,\labell{sstring333}\eeqa

The entire amplitude can be divided out to two separate correlation functions. We also need to make use of the generalized Wick-like rule.
Define  
\beqa
s&=&\frac{-\alpha'}{2}(k_1+k_3)^2,\quad t=\frac{-\alpha'}{2}(k_1+k_2)^2,\quad u=\frac{-\alpha'}{2}(k_2+k_3)^2
\nonumber\eeqa 
 It is easy to find out all the correlators and show that the amplitude is SL(2,R) invariant . Note that we did gauge fixing as $x_1=0,x_2=1,x_3=\infty$  so that the amplitude after gauge fixing is written down as follows 
\beqa
{\cal A}^{C^{-2}A^{0} T^{0} T^{0}}&\sim&\int dx_{4}dx_{5}(P_{-}\fsC_{(n-1)}M_p)^{\al\be} \xi_{1a}
(-4 k_{2b}k_{3c})x_{45}^{-2(t+s+u+1)}|x_{4}|^{2t+2s+1}|1-x_4|^{2t+2u-1/2}\nonumber\\&&\times
\bigg[\bigg((\Gamma^{cb}C^{-1})_{\alpha\beta}+2\eta^{bc}(C^{-1})_{\alpha\beta}(\frac{Re[1-x_4]}{x_{45}})\bigg) I_{11}+ik_{1d}|x_{4}|^{-2} x_{45} I_{22}\bigg]\Tr(\lam_1\lam_2\lam_3)
\nonumber\eeqa
where
\beqa
I_{11}&=&-i |x_{4}|^{-2}(x_4+x_5)(k^{a}_{2}+k^{a}_{3})+2i k^{a}_{2} ,\nonumber\\
I_{22}&=&\bigg[(\Gamma^{cbad}C^{-1})_{\alpha\beta}  +2\bigg(\eta^{db}(\Gamma^{ca}C^{-1})_{\alpha\beta}-\eta^{ab}(\Gamma^{cd}C^{-1})_{\alpha\beta}\bigg)(\frac{x-|x_{4}|^{2}}{x_{45}})\nonumber\\&&
+2\bigg(-\eta^{dc}(\Gamma^{ba}C^{-1})_{\alpha\beta} +\eta^{ac}(\Gamma^{bd}C^{-1})_{\alpha\beta}\bigg) \frac{x}{x_{45}}\nonumber\\&&
+2\bigg(\eta^{bc}(\Gamma^{ad}C^{-1})_{\alpha\beta} \bigg)(\frac{1-x}{x_{45}})
+4 (-\eta^{db}\eta^{ac}+\eta^{dc}\eta^{ab})(\frac{x^2-x|x_{4}|^2}{(x_{45})^2})\bigg]
\nonumber\eeqa

All the integrations on upper half plane can be evaluated on the location of closed string and the final result can  be written  down just in terms of  four Gamma functions over three Gamma functions by using the important results of the following integrals
\beqa
 \int d^2 \!z |1-z|^{a} |z|^{b} (z - \bar{z})^{c}
(z + \bar{z})^{d}\nonumber
\eeqa
that are accommodated for $d=0,1$ in \cite{Fotopoulos:2001pt} and for $d=2$  in \cite{Hatefi:2012wj} where
$z=x_4=x+iy, \bar z=x_5=x-iy, x_{ij}=x_i-x_j$.

First of all let us see what happens to the terms that are apparently absent in symmetric picture. Hence, we are now considering the terms 
in asymmetric amplitude that after all just involve $\Tr(P_{-}\fsC_{(n-1)}M_p)$ as follows:

\beqa {\cal A}_{3}^{C^{-2} A^{0} T^{0} T^{0} }&=& -4i I_{33} \xi_{1a} (2)^{-2(t+s+u)-3}\pi{\frac{\Gamma(-u)
\Gamma(-s-\frac{3}{4})\Gamma(-t+\frac{1}{4})\Gamma(-t-s-u-1)}
{\Gamma(-u-t+\frac{1}{4})\Gamma(-t-s+\frac{1}{2})\Gamma(-s-u+\frac{1}{4})}}\nonumber \eeqa

where 

\beqa
I_{33}&=&2(u+\frac{1}{2})(k^{a}_{2}+k^{a}_{3})(-t+\frac{1}{4})(-s-u-\frac{3}{4})\nonumber\\&&
-2(u+\frac{1}{2})(k^{a}_{2})(-t-s-\frac{1}{2})(-s-u-\frac{3}{4})\nonumber\\&&
-2(u+\frac{1}{2})(k^{a}_{2}+k^{a}_{3})(\frac{1}{2}(-s-\frac{3}{4})-u(-t+\frac{1}{4}))\nonumber\\&&
+2(u+\frac{1}{2})(k^{a}_{2})(-t-s-\frac{1}{2})(-u)\nonumber\\&&
+(\frac{1}{2}(-s-\frac{3}{4})-u(-t+\frac{1}{4}))(2(t+\frac{1}{4})k^{a}_{3}-2(s+\frac{1}{4})k^{a}_{2})\nonumber\\&&
+u(-t-s-\frac{1}{2})(2(t+\frac{1}{4})k^{a}_{3}-2(s+\frac{1}{4})k^{a}_{2})
\eeqa

One might think that these terms are needed in the entire S-matrix, however, after some simplification, one can readily see that the sum of  coefficients of $k^{a}_{2}$ and $k^{a}_{3}$ separately vanishes  so that the whole ${\cal A}_{3}^{C^{-2} A^{0} T^{0} T^{0} }$ vanishes and has no contribution to the S-matrix at all.

 Eventually after having performed massive computations on all the integrations and making use of momentum conservation along the world volume of brane, one is able to get the final form of S-matrix in asymmetric picture as follows

\beqa {\cal A}^{C^{-2} A^{0} T^{0} T^{0} }&=&{\cal A}_{1}+{\cal A}_{2}\labell{711u}\eeqa
where
\beqa
{\cal A}_{1}&\!\!\!\sim\!\!\!&-4i \xi_{1a} k_{1d}k_{2b}k_{3c}\Tr(P_{-}\fsC_{(n-1)}M_p\Gamma^{cbad})
L_1,
\nonumber\\
{\cal A}_{2}&\sim&-4i  p_{d}\Tr(P_{-}\fsC_{(n-1)}M_p\Gamma^{ad})\bigg(2k_2.\xi_1 k^{a}_{3}(s+\frac{1}{4}) +2k_3.\xi_1 k^{a}_{2}(t+\frac{1}{4}) -\xi_{1a}(t+\frac{1}{4})(s+\frac{1}{4})\bigg)L_2
\nonumber\eeqa
with \beqa
L_1&=&(2)^{-2(t+s+u)-1}\pi{\frac{\Gamma(-u)
\Gamma(-s+\frac{1}{4})\Gamma(-t+\frac{1}{4})\Gamma(-t-s-u)}
{\Gamma(-u-t+\frac{1}{4})\Gamma(-t-s+\frac{1}{2})\Gamma(-s-u+\frac{1}{4})}},\nonumber\\
L_2&=&(2)^{-2(t+s+u+1)}\pi{\frac{\Gamma(-u+\frac{1}{2})
\Gamma(-s-\frac{1}{4})\Gamma(-t-\frac{1}{4})\Gamma(-t-s-u-\frac{1}{2})}
{\Gamma(-u-t+\frac{1}{4})\Gamma(-t-s+\frac{1}{2})\Gamma(-s-u+\frac{1}{4})}},\nonumber\label{Ls}
\eeqa

 Just for the record we write down the results in all different symmetric picture (in terms of RR's field strength not its potential any more) as follows:
 
 \beqa
{\cal A}^{C^{-1} A^{0} T^{-1} T^{0} }&=&{\cal A}^{C^{-1} A^{-1} T^{0} T^{0} }=\frac{i\mu_p}{2\sqrt{\pi}}\left[k_{3c} k_{2b} \xi_{1a}\Tr\bigg((P_{-}\fsH_{(n)}M_p \Gamma^{cba})
\bigg) L_1
+\Tr\bigg((P_{-}\fsH_{(n)}M_p)
\ga^{a}\bigg) L_2\right. \nonumber\\&&\left.\times
\bigg\{k_{2a}(t+1/4)(2\xi_1.k_{3})
+k_{3a}(s+1/4)(2\xi_1.k_{2})-\xi_{1a}(s+1/4)(t+1/4)\bigg\}\right]\labell{gen}\eeqa

Now if we use momentum conservation along the world volume of brane  as 
$(k_1+k_2+k_3)^d=-p^d$ and  use the fact that ${\cal A}_{1}$ of \reef{711u} is symmetric with respect to $k_2$,$k_3$ but antisymmetric under $\epsilon^{a_0..a{p-3}bcd}$, therefore the first term of asymmetric picture does match with the 1st term of symmetric amplitude and likewise for the other terms, where
$p_d \fsC_{(n-1)}= \fsH_{(n)}$ has been used.

\vskip.2in

This evidently confirms that the vertex operator of RR in asymmetric picture is exact and complete. It also confirms that no extra potential terms needed to be added to RR potential vertex operator  in asymmetric picture. It  also reveals that for the
mixed world volume S-matrices of closed string RR, gauge fields and tachyons, even in five and higher point functions there is no picture dependence at all. However,  the story gets complicated for the higher point functions of the mixed closed string RR and scalar fields as we will point out in detail in the next section.

\section{ All order Bulk Singularities of $<C^{(-2)}\phi^{(0)} T^{(0)} T^{(0)}>$ S-matrix in Brane Anti Brane System }

The general structures of vertex operators with their Chan-Paton  factors in brane anti brane system can be shown by

 \beqa
V_{\phi}^{(0)}(x) &=& \xi_{1i}(\partial X^i(x)+i\alpha'k.\psi\psi^i(x))e^{\alpha'ik.X(x)}\otimes I\nonumber\\
V_\phi^{(-1)}(x)&=&e^{-\phi(x)}\xi_i\psi^i(x)e^{ \alpha'iq\inn X(x)}\lam\otimes \sigma_3 \nonumber\\
V_\phi^{(-2)}(x)&=& e^{-2\phi(x)}V_{\phi}^{(0)}(x)\nonumber\\
V_{C}^{(-\frac{1}{2},-\frac{1}{2})}(z,\bar{z})&=&(P_{-}\fsH_{(n)}M_p)^{\alpha\beta}e^{-\phi(z)/2}
S_{\al}(z)e^{i\frac{\alpha'}{2}p\cd X(z)}e^{-\phi(\bar{z})/2} S_{\be}(\bar{z})
e^{i\frac{\alpha'}{2}p\cd D \cd X(\bar{z})}\lam\otimes \sigma_3 \nonumber\eeqa

The ultimate form of a field strength RR , one transverse scalar field and two real tachyons in the following picture is derived in
\cite{Hatefi:2012cp} to be 

 \beqa {\cal A}^{C^{(-1)}\phi^{(-1)} T^{(0)} T^{(0)}}&=&{\cal A}_{1}+{\cal A}_{2}\labell{181u}\eeqa
where
\beqa
{\cal A}_{1}&\!\!\!\sim\!\!\!&-8\xi_{1i}k_{2a}k_{3b} 2^{-3/2}
\Tr(P_{-}\fsH_{(n)}M_p\Gamma^{bai}),
\nonumber\\
{\cal A}_{2}&\sim&8\xi_{1i} 2^{-3/2}\Tr(P_{-}\fsH_{(n)}M_p \gamma^{i}) (t+\frac{1}{4}) (s+\frac{1}{4}) L_2
\labell{4859}\eeqa
where

 $L_1,L_2$ are already given in the last section. 
 
 Note that since there is a non zero correlation function of $<e^{ip.X(z)} (\partial X^i(x_1))>$ and more importantly, there is no Ward identity for mixed closed RR and transverse scalar fields, 
 one needs to think of all the bulk singularities  that do include an infinite number of  $p.\xi$ terms  as well as keep an eye on asymmetric picture for which we are going to illustrate on that right now. Note that in asymmetric amplitude,  we now get some extra singularities in the bulk of brane anti brane system that  carry  $p.\xi$ terms. 
 
\vskip.3in

Indeed by dealing with the same elements and coming over  its last symmetric change of picture or for the other symmetric result of  $< C^{(-1)}  \phi^{(0)}  T^{(-1)} T^{(0)} >$  we get the final answer 
 in the presence of brane anti brane as below

\beqa {\cal A}^{C^{(-1)}  \phi^{(0)}  T^{(-1)} T^{(0)}}&=&{\cal A}_{1}+{\cal A}_{2}+{\cal A}_{3}\labell{181u}\eeqa

so that this turn their part gets replaced by 
\beqa
{\cal A}_{1}&\!\!\!\sim\!\!\!&- 2^{3/2} i p^i\xi_{1i}k_{3b}
\Tr(P_{-}\fsH_{(n)}M_p\gamma^{b})L_1,
\nonumber\\
{\cal A}_{2}&\sim&- 2^{3/2} i\xi_{1i} \bigg\{\Tr(P_{-}\fsH_{(n)}M_p \Gamma^{bia})\bigg\}k_{1a} k_{3b} L_1
\nonumber\\
{\cal A}_{3}&\sim&- 2^{3/2} i\xi_{1i} \bigg\{\Tr(P_{-}\fsH_{(n)}M_p \gamma^{i})\bigg\} (t+\frac{1}{4}) (s+\frac{1}{4}) L_2
\labell{4k8}\eeqa

Note that one might make use of momentum conservation along the world volume of brane and use some sort of generalized Bianchi 
identities \footnote{
\beqa
\xi_{1i} k_{3b}\bigg(-p_a\eps^{a_{0}\cdots a_{p-2}ab}H^{i}_{a_{0}\cdots a_{p-2}}+p^i\eps^{a_{0}\cdots a_{p-1}b}H_{a_{0}\cdots a_{p-1}}\bigg)=0 \label{oo}\eeqa}
 to get rid of the first term of \reef{4k8} , however, this did happen if and only if we would have not been able to produce those infinite u-channel Bulk singularities that carry $p.\xi$ term in an Effective Field Theory. 
 Let us further elaborate on that by taking into account the asymmetric result as well.
  

\vskip.2in

Indeed it  is easy to find out all the correlators of a C-field in asymmetric picture and a real transverse scalar field and two real tachyons of brane anti brane system of  $ <C^{(-2)}  \phi^{(0)}  T^{(0)} T^{(0)}> $
and show that the amplitude is SL(2,R) invariant.  Note that we did gauge fixing as $x_1=0,x_2=1,x_3=\infty$  so that the amplitude after gauge fixing is written down as follows 
\beqa
{\cal A}^{C^{-2} \phi^{0} T^{0} T^{0}}&\sim&\int dx_{4}dx_{5}(P_{-}\fsC_{(n-1)}M_p)^{\al\be} \xi_{1i}
(-4 k_{2b}k_{3c})x_{45}^{-2(t+s+u)-1}|x_{4}|^{2t+2s-1}|1-x_4|^{2t+2u-1/2}\nonumber\\&&\times
\bigg[-i p^i\bigg((\Gamma^{cb}C^{-1})_{\alpha\beta}+2\eta^{bc}(C^{-1})_{\alpha\beta}(\frac{Re[1-x_4]}{x_{45}})\bigg) +ik_{1a}
   I_{44}\bigg]\Tr(\lam_1\lam_2\lam_3)
\nonumber\eeqa
where
\beqa
I_{44}&=&\bigg[(\Gamma^{cbia}C^{-1})_{\alpha\beta}  +2\bigg(\eta^{ab}(\Gamma^{ci}C^{-1})_{\alpha\beta}\bigg)(\frac{x-|x_{4}|^{2}}{x_{45}})\nonumber\\&&
-2\bigg(\eta^{ac}(\Gamma^{bi}C^{-1})_{\alpha\beta} \bigg) \frac{x}{x_{45}}
+2\bigg(\eta^{bc}(\Gamma^{ia}C^{-1})_{\alpha\beta} \bigg)(\frac{1-x}{x_{45}})\bigg]
\nonumber\eeqa

Taking the integrations on upper half plane and doing more simplifications, one gets the essential form of the amplitude , where more ingredients can be found from \cite{Hatefi:2015gwa} so that

\beqa {\cal A}^{C^{-2}\phi^{0} T^{0} T^{0}}&=&{\cal A}_{1}+{\cal A}_{2}+{\cal A}_{3}+{\cal A}_{4}\labell{1m81u}\eeqa
where
\beqa
{\cal A}_{1}&\!\!\!\sim\!\!\!&ip^i\xi_{1i}(4 k_{2b}k_{3c})
\Tr(P_{-}\fsC_{(n-1)}M_p\Gamma^{cb})L_1,
\nonumber\\
{\cal A}_{2}&\sim&4i p^i \xi_{1i} \bigg\{\Tr(P_{-}\fsC_{(n-1)}M_p)\bigg\} (t+\frac{1}{4}) (s+\frac{1}{4}) L_2
\nonumber\\
{\cal A}_{3}&\sim&-4i \xi_{1i} k_{1a}k_{2b}k_{3c} \bigg\{\Tr(P_{-}\fsC_{(n-1)}M_p \Gamma^{cbia})\bigg\} L_1
\nonumber\\
{\cal A}_{4}&\sim&4i  \xi_{1i} \bigg\{\Tr(P_{-}\fsC_{(n-1)}M_p\Gamma^{bi})\bigg\}  (t+\frac{1}{4}) (s+\frac{1}{4}) L_2 (k_{1b}+k_{2b}+k_{3b})
\labell{mm2uu}\eeqa

notice that in this asymmetric result for mixed closed string RR in the presence of two real tachyons of brane anti brane and a transverse scalar field we have got four terms where two of them ,namely, 
${\cal A}_{1}$  and ${\cal A}_{2}$ are extra bulk singularities that have just been shown up in asymmetric result as have already been pointed out. These terms carry the scalar 
product of RR momentum and scalar polarization which 
 are both located in the bulk , that is why we call them bulk singularity structures and in fact $L_1$ does have infinite u- channel singularities where as $L_2$ carries an infinite number of $(t'+s'+u)$ channel singularities where $s'=s+1/4, t'=t+1/4$ have been defined.

 \vskip.1in

 All $(p.\xi)$'s terms do have  potentially several things to do  with worthwhile issues within the field of  perturbative string theory that has been going on over last few years and started by series of papers  by Witten \cite{Witten:2012bh},
 also bulk terms can have different meanings if one starts to think about  still non trivial moduli spaces \cite{Witten:2012bg}.

\vskip.1in

One might consider the momentum conservation
$(k_1+k_2+k_3)^a=-p^a$ to actually apply the Bianchi identity that has been used for BPS amplitudes  ( $p^b\eps^{a_{0}\cdots a_{p-1}b}=0$)  to this brane anti brane S-matrix , however, we claim that not all the Bianchi identities of BPS cases can hold for non-BPS, 
nor for brane anti brane amplitudes. The reasons for that conclusion is as follows.  First of all we do need to keep track of those bulk singularities in the S-Matrix, secondly, not all  the equations of  BPS cases can be true  for non-BPS or  non-supersymmetric S-Matrices ,
indeed all the equations of BPS cases are so manifest, meanwhile this might not occur to non-BPS cases as things get changed after symmetry breaking happened. Hence for non-BPS branes, one needs to break certain several equations of BPS branes.

\vskip.3in

We already knew from EFT that we must have an infinite number of u-channel gauge field  poles, therefore we no longer can add up the 3rd term of \reef{mm2uu} with the 1st term of \reef{mm2uu} to get generalized Bianchi identity. Indeed the same 
story  holds for the other terms in the above amplitude.  Basically  all the terms appearing in this amplitude, are needed in an EFT formalism as we start producing all those new couplings in the next section.

 
 \section{ All order u- channel Bulk Singularity Structures of $<C^{-2} \phi^0 T^0 T^0>$ }
 
 It has been comprehensively explained in  \cite{Hatefi:2012cp} how to explore the momentum expansions of tachyons, nevertheless we just hint that , by applying momentum conservation 
 we get  the constraint as $s'+t'+u=-p^ap_a $  where $s'=s +\frac{1}{4} , t'=t +\frac{1}{4} $ and as highlighted in the last section for brane anti-brane system the soft condition $p^a p_a \rightarrow 0 $ holds, thus 
 the expansion is $u \rightarrow 0 , s\rightarrow \frac{-1}{4}, t \rightarrow \frac{-1}{4}$.  In order to deal with  u-channel gauge poles, we need to first employ   ${\cal A}_{1}$ and  eventually ${\cal A}_{3}$ , extract the traces and consider all order $\alpha'$ expansion of $L_1$  as below
 \beqa
 ({\cal A}_{1})^{C^{-2} \phi^0 T^0 T^0 }&=&  \eps^{a_{0}\cdots
a_{p-2}cb} p^i C_{a_{0}\cdots a_{p-2}}  \frac{64}{(p-1)!}L_1\xi_{1i}k_{3c}k_{2b}\labell{pnbb}\eeqa
  
 Indeed one can see that some  terms of  $L_1$ expansion have the same behaviour as we have seen in the case of $<C^{-2} T^0 T^0>$ 
 S-matrix , as 
 \beqa
L_1&=&
\pi^{3/2}\bigg(\frac{-1}{u}
+4\ln(2)+\bigg(\frac{\pi^2}{6}-8\ln(2)^2\bigg)u-\frac{\pi^2}{6}\frac{(s'+t')^2}{u}+\cdots\bigg)\labell{line} \eeqa
 
 Given the fact that there are prescriptions that one can investigate all order $\alpha'$ corrections of string amplitudes, we just write down the compact and all order expansion of $L_1$  that has been achieved in  \cite{Hatefi:2012cp}

 \beqa
L_1&=&
\pi^{3/2}\left(-\frac{1}{u}\sum_{n=-1}^{\infty}b_n(s'+t')^{n+1}+\sum_{p,n,m=0}^{\infty}f_{p,n,m}u^p\left(s't'\right)^n(s'+t')^m\right)
\label{ee1}\eeqa
 with some of its own coefficients 
   \beqa
&&b_{-1}=1,\,b_0=0,\,b_1=\frac{\pi^2}{6},\,b_2=2\z(3),a_0=4ln2,\nonumber\\&&
a_1=\frac{\pi^2}{6}-8ln(2)^2,a_2=\frac{2}{3}(-\pi^2 ln2+3\z(3)+16ln(2)^3),\nonumber\\
&&f_{0,0,2}=\frac{2}{3}\pi^2\ln(2),\,f_{0,1,0}=-14\z(3),
f_{0,0,3}=8\z(3)\ln(2),\nonumber\\
\eeqa
 
In order to actually produce all infinite u- channel bulk singularities,  one first needs to consider the following Feynman rule 
\beqa {\cal
A}&=&V_a(C_{p-1},\phi,A)G_{ab}(A)V_b(A,T_1,T_2)\labell{amp37}\eeqa
with the following EFT  vertex and propagator

 \beqa G_{ab}(A) &=&\frac{ i\delta_{ab}}{(2\pi\alpha')^2 T_p
k^2}\nonumber\\
V_b(A,T_1,T_2)&=&T_p(2\pi\alpha')(k_{2b}-k_{3b})\label{aa1} \eeqa

where $k^2=\frac{\alpha'}{2}(k_2+k_3)^2=-u$ and the kinetic term of gauge field  $\Tr(\frac{-1}{4} F_{ab}F^{ba})$ has been taken. 
  $V_b(A,T_1,T_2)$ is derived from kinetic term of Tachyons in DBI action $\Tr(2\pi\alpha' D_aT D^aT)$. 
 Notice that all kinetic terms of gauge  field, scalar and tachyons have been already fixed, so that they do not receive any correction terms at all.
 
Now consider the mixing
Chern-Simons coupling as well as Taylor expanded of scalar field to get to 
 $V_a(C_{p-1},\phi,A)$ as follows
 \beqa
 \mu_p (2\pi\alpha')^2\int_{\sum_{(p+1)}}\Tr\bigg(\partial_i C_{p-1}\wedge F \phi^i\bigg)\labell{newrr}
 \eeqa

Particular attention should be drawn to the fact that  we do not need to take integration by parts, having set \reef{newrr}, we get to

\beqa
V_a(C_{p-1},\phi,A
)&=&\mu_p(2\pi\alpha')^2\frac{1}{(p-1)!}\epsilon^{a_0\cdots a_{p-2}
} p^i C_{a_0\cdots
a_{p-2}} k_{a}\xi_{1i}\label{esi44}\eeqa

where $k$ is  the momentum of off-shell gauge field, $k_a=-(k_2+k_3)_a$.
If we do normalise the string amplitude by $\frac{i\mu_p}{8\sqrt{\pi}}$ and replace \reef{esi44} and \reef{aa1} inside \reef{amp37} then we are precisely able to produce  the first u-channel gauge pole of \reef{pnbb} and \reef{line}  
in an Effective field theory as well.

\vskip .1in

Notice that the second and third terms of \reef{line} do relate to contact terms and both can be produced in EFT by taking the following couplings accordingly. \footnote{ 

\beqa
\frac{i}{2} \mu_p \beta^2 (2\pi\alpha')^3  \Tr\bigg(\partial_{i} C_{p-1}\wedge DT \wedge DT^{*}(\phi^1+\phi^2)^i\bigg)
\nonumber\\
\frac{i}{2} \mu_p  (2\pi\alpha') (\alpha')^2 \bigg(\frac{\pi^2}{6}-8 ln2^2\bigg) \Tr\bigg(\partial_{i} C_{p-1}\wedge D^aD_a( DT \wedge DT^{*})(\phi^1+\phi^2)^i\bigg)
\label{nnew22}
\eeqa.}

\vskip .2in

As we can see from the expansion of $L_1$, it has so many singularities and in order to produce them, one needs to apply  proper all order $\alpha'$ corrections to Chern-Simons coupling , because the kinetic terms of gauge field and tachyons do not gain any correction as they have been fixed in DBI action. Let us apply all order $\alpha'$ higher derivative corrections as \beqa
 \mu_p (2\pi\alpha')^2\int_{\sum_{(p+1)}}\partial_iC_{p-1}\wedge  \sum_{n=-1}^{\infty}b_{n}(\alpha')^{n+1} \Tr\bigg(
   D_{a_{1}}\cdots D_{a_{n+1}}F D^{a_{1}}\cdots D^{a_{n+1}} \phi^i\bigg)\labell{new287}
 \eeqa
to be able to obtain  all order  vertex of
$V_a(C_{p-1},\phi,A)$ as follows

 \beqa
V_a(C_{p-1},\phi,A)&=&\mu_p(2\pi\alpha')^2\frac{1}{(p-1)!}\epsilon^{a_0\cdots a_{p-2}
} p^i C_{a_0\cdots
a_{p-2}}(k_2+k_3)^{a}\xi_{i}\sum_{n=-1}^{\infty}b_n(\alpha'k_1\cdot
k)^{n+1}\label{7ui}\eeqa

substituting  \reef{7ui}  and \reef{aa1} inside \reef{amp37}, we are precisely able to produce  all order u-channel gauge poles of \reef{pnbb}   in an Effective field theory as below
\beqa {\cal
A}=\mu_p(2\pi\alpha')\frac{2i}{(p-1)!u}\epsilon^{a_0\cdots
a_{p-2}bc} p^i C_{a_0\cdots
a_{p-2}}k_{2b}k_{3c}\xi_{1i}\sum_{n=-1}^{\infty}b_n\left(\frac{\alpha'}{2}\right)^{n+1}(s'+t')^{n+1}\labell{AAbb}\eeqa

Note that all order contact interactions of this S-matrix can be found in section 3.1 of \cite{Hatefi:2012cp}. Let us produce  
another all infinite bulk singularities of  ${\cal A}_{3}$. Thus we extract the traces and consider all order $\alpha'$ expansion of $L_1$  as below
 \beqa
  ({\cal A}_{3})^{C^{-2} \phi^0 T^0 T^0 }&=&  -\eps^{a_{0}\cdots
a_{p-3}abc} p_a C^{i}_{a_{0}\cdots a_{p-3}} \frac{64}{(p-1)!}L_1\xi_{1i}k_{3c}k_{2b}\labell{pnbb22}\eeqa
 where we have used momentum conservation as well as the fact that the 3rd term of amplitude was symmetric under interchanging both $k_2, k_3$ but  also is  anti symmetric as it does involve  $\eps$ tensor , therefore
 just  $p^a$ terms remain after applying momentum conservation to the 3rd term of S-matrix.

In order to actually produce  the second kind of an infinite u- channel bulk singularities of ${\cal A}_{3}$, one needs to consider the  same Feynman rule as appeared in \reef{amp37} where the same definitions for propagator and $V_b(A,T_1,T_2)$ kept held. But one has to define a new sort of coupling as follows 
 
 \beqa
\frac{1}{(p-1)!} \mu_p (2\pi\alpha')^2\int_{\sum_{(p+1)}}\Tr\bigg( C^{i}_{a_0...a_{p-3}} F_{a_{p-2}a_{p-1}} D_{a_{p}}\phi^i\bigg) \epsilon^{a_0...a_p}\labell{newrr22}
 \eeqa

where in \reef{newrr22} the scalar field comes from pull-back of brane and the Chern-Simons coupling also has been taken into account. As we discussed there are infinite singularities and in order to reconstruct them out, one needs to apply all order $\alpha'$ higher derivative corrections to the new coupling  \reef{newrr22} as follows

\beqa 
\frac{\mu_p (2\pi\alpha')^2}{(p-1)!}\int_{\sum_{(p+1)}}\Tr\bigg( C^{i}_{a_0...a_{p-3}} \sum_{n=-1}^{\infty}b_{n}(\alpha')^{n+1}  D_{a_{1}}\cdots D_{a_{n+1}} F_{a_{p-2}a_{p-1}} D^{a_{1}}\cdots D^{a_{n+1}} D_{a_{p}}\phi^i\bigg)\labell{new28722}
 \eeqa


to be able to obtain all order  vertex of
$V_a(C_{p-1},\phi,A)$ as follows

 \beqa
V_a(C_{p-1},\phi,A)&=&\mu_p(2\pi\alpha')^2\frac{1}{(p-1)!}\epsilon^{a_0\cdots a_{p-1}
} p_{a_{p-1}} C^{i}_{a_0\cdots
a_{p-3}}(k_2+k_3)_{a_{p-2}}\xi_{i}\nonumber\\&&\times\sum_{n=-1}^{\infty}b_n(\alpha'k_1\cdot
k)^{n+1}\label{7ui22}\eeqa
Notice to the point that in \reef{newrr22}, the derivative $D_{a_{p}}$ can not act on $d_{a_{p-2}} A_{a_{p-1}}$ because it is symmetric under interchanging the derivatives but is antisymmetric under $\epsilon$ tensor so the result is zero , hence after taking integration by parts $D_{a_{p}}$ can just act on $C^{i}_{a_0...a_{p-3}}$ .

Replacing \reef{7ui22}  and \reef{aa1} inside \reef{amp37}, we are precisely able to produce  all order u-channel gauge poles of \reef{pnbb22}   in an effective field theory as below
\beqa {\cal
A}=\mu_p(2\pi\alpha')\frac{2i}{(p-1)!u}\epsilon^{a_0\cdots
a_{p-2}abc} p^a C^{i}_{a_0\cdots
a_{p-3}}k_{2b}k_{3c}\xi_{1i}\sum_{n=-1}^{\infty}b_n\left(\frac{\alpha'}{2}\right)^{n+1}(s'+t')^{n+1}\labell{AAbb}\eeqa

Therefore by making use of the new coupling  \reef{newrr22} we were able to exactly regenerate all infinite u-channel bulk singularities of this S-matrix.

\section {Infinite $(u+s'+t')$ channel Bulk singularities }

The amplitude is non zero for $n-1=p+1$ case  and it involves both ${\cal A}_{2}$ and ${\cal A}_{4}$  terms, also note that the amplitude for this case is symmetric under interchanging the tachyons, let us extract the traces and write them down explicitly as below

\beqa
{\cal A}_{2}&=&\frac{8i \mu_p}{\sqrt{\pi}(p+1)!} p^i \xi_{1i}   \eps^{a_{0}\cdots a_{p}} C_{a_{0}\cdots a_{p}} t's' L_2
\nonumber\\
{\cal A}_{4}&=&\frac{8i \mu_p}{\sqrt{\pi}(p+1)!} \xi_{1i} \eps^{a_{0}\cdots a_{p-1}b} C^{i}_{a_{0}\cdots a_{p-1}} t's' L_2 (k_{1b}+k_{2b}+k_{3b})
\labell{mm2}\eeqa

Note that there is no Ward identity associated to the scalar field in the presence of RR.  The closed form of the expansion of $ t' s' L_2$  is given by
\beqa
 t' s' L_2&=&\frac{\sqrt{\pi}}{2}\bigg(\frac{-1}{(t'+s'+u)}+ \sum_{n=0}^{\infty}a_n(s'+t'+u)^n+\frac{\sum_{n,m=0}^{\infty}l_{n,m}(s'+t')^n(t's')^{m+1}}{(t'+s'+u)}\nonumber\\&&+\sum_{p,n,m=0}^{\infty}e_{p,n,m}(s'+t'+u)^p(s'+t')^n(t's')^{m+1}\bigg)\label{mmn}\eeqa

$l_{n,m}$ and $e_{p,n,m}$ are
 \beqa
l_{0,0}=-\pi^2/3,\,&&l_{1,0}=8\z(3)\\
l_{2,0}=-7\pi^4/45,\, l_{0,1}=\pi^4/45,\,&&\,l_{3,0}=32\z(5),\, l_{1,1}=-32\z(5)+8\z(3)\pi^2/3\nonumber\\
e_{0,0,0}=\frac{2}{3}\left(2\pi^2\ln(2)-21\z(3)\right),&& e_{1,0,0}=\frac{1}{9}\left(4\pi^4-504\z(3)\ln(2)+24\pi^2\ln(2)^2\right)\nonumber
\eeqa

Let us   consider  all order singularities related to ${\cal A}_{4}$ term , where we have applied momentum conservation along the world volume of brane and extracted the trace. We also take all the singularities in the expansion of $t's' L_2$  and replace them in ${\cal A}_{4}$ term so that all poles are written down as follows

\beqa
{\cal A}_{4}&=&-\frac{4i \mu_p}{(p+1)!} \xi_{1i} \eps^{a_{0}\cdots a_{p-1}b} p^b C^{i}_{a_{0}\cdots a_{p-1}}  \frac{\sum_{n,m=0}^{\infty}l_{n,m}(s'+t')^n(t's')^{m+1}}{(t'+s'+u)} \labell{mm222}\eeqa

The  Lagrangian for two scalar, two tachyon couplings for non-BPS brane was defined as
\beqa
L (\phi,\phi,T,T)&=& -2T_p(\pi\alpha')^3 S\Tr \bigg(
m^2 T^2(D_a\phi^iD^a\phi_i)+\frac{\alpha'}{2}D^{\alpha} T D_{\alpha} T D_a\phi^iD^a\phi_i \nonumber\\&&-
\alpha' D^{b} T D^{a} T D_a\phi^iD_b\phi_i \bigg)\labell{dbicoupling} \eeqa

Note that  all order corrections to two tachyons and two scalar fields of brane anti brane system have been derived in  \cite{Hatefi:2012cp}, but for the completeness we just write them down  because in order to produce an infinite singularities , 
we  need to deal with them as well. 

 \beqa
L&=&-2T_p(\pi\alpha')(\alpha')^{2+n+m}\sum_{n,m=0}^{\infty}( L_{1}^{nm}+ L_{2}^{nm}+ L_{3}^{nm}+ L^{nm}_{4}),\labell{lagrango212}\eeqa
where
\beqa
L_1^{nm}&=&m^2
\Tr\left(\frac{}{}a_{n,m}[ D_{nm}(T T^* D_a\phi^{(1)i}D^a\phi^{(1)}_i)+  D_{nm}( D_a\phi^{(1)i}D^a\phi^{(1)}_i  T T^*)+h.c.]\right.\nonumber\\
&&\left.-\frac{}{}b_{n,m}[ D'_{nm}( T D_a\phi^{(2)i} T^* D^a\phi^{(1)}_i)+ D'_{nm}( D_a\phi^{(1)i} T D^a\phi^{(2)}_i T^*)+h.c.]\right),\nonumber\\
L_2^{nm}&=&\Tr\left(\frac{}{}a_{n,m}[ D_{nm}(D^{\alpha} T D_{\alpha} T^* D_a\phi^{(1)i}D^a\phi^{(1)}_i)+ D_{nm}( D_a\phi^{(1)i}D^a\phi^{(1)}_i D^{\alpha} T D_{\alpha} T^*)+h.c.]\right.\nonumber\\
&&\left.-\frac{}{}b_{n,m}[ D'_{nm}(D^{\alpha}  T D_a\phi^{(2)i} D_{\alpha} T^* D^a\phi^{(1)}_i)+ D'_{nm}( D_a\phi^{(1)i} D_{\alpha} T D_a\phi^{(2)}_i D^{\alpha} T^*)+h.c.]\right),\nonumber\\
L_3^{nm}&=&-\Tr
\left(\frac{}{}a_{n,m}[ D_{nm}(D^{\beta} T D_{\mu} T^* D^\mu\phi^{(1)i} D_\beta\phi^{(1)}_i)+ D_{nm}( D^\mu\phi^{(1)i} D_\beta \phi^{(1)}_iD^{\beta} T D_{\mu} T^*)+h.c.]\right.\nonumber\\
&&\left.-\frac{}{}b_{n,m}[ D'_{nm}(D^{\beta} T D^\mu\phi^{(2)i}D_{\mu} T^* D_\beta\phi^{(1)}_i)+ D'_{nm}(D^\mu\phi^{(1)i} D_{\mu} T  D_\beta\phi^{(2)}_i D^{\beta}T^*)+h.c.]\right),\nonumber\\
L_4^{nm}&=&-\Tr\left(\frac{}{}a_{n,m}[ D_{nm}(D^{\beta} T D^{\mu} T^* D_\beta\phi^{(1)i}D_\mu\phi^{(1)}_i)
+ D_{nm}( D^\beta\phi^{(1)i}D^\mu\phi^{(1)}_iD_{\beta} T D_{\mu} T^*)+h.c.]\right.\nonumber\\
&&\left.-\frac{}{}b_{n,m}[ D'_{nm}(D^{\beta} T D_\beta\phi^{(2)i}D^{\mu} T^* D_\mu\phi^{(1)}_i)+ D'_{nm}( D_\beta\phi^{(1)i} D_{\mu} T  D^\mu\phi^{(2)}_iD^{\beta} T^*)+h.c.]
\right)\label{hdts212}
\eeqa

One might read off  all the definitions of   $D_{nm}(EFGH), D'_{nm}(EFGH)$ from \cite{Hatefi:2012wj}. In order to actually produce all infinite singularities $(u+s'+t')$ , one needs to employ the following rule

\beqa {\cal
A}&=&V_{\alpha}^{i}(C_{p+1},\phi)G_{\alpha\beta}^{ij}(\phi)V_{\beta}^{j}(\phi,\phi_1,
T_2,T_2),\labell{amp5419} \eeqa
where the scalar propagator $G_{\alpha\beta}^{ij}(\phi)=\frac{-i\delta_{\alpha\beta}\delta^{ij}}{T_p(2\pi\alpha')^2 (t'+s'+u)}$ can be derived from the scalar field's kinetic term that has been fixed in  DBI action. 

Let us introduce new coupling as 
\beqa
(2\pi\alpha')\mu_p\frac{1}{(p+1)!} \int_{\Sigma_{p+1}}\eps^{a_0\cdots
a_{p}}
 C^{i }_{a_0\cdots a_{p-1}} D_{a_{p}} \phi^i \label{esi897}\eeqa

where in the above new coupling,  the scalar field has been taken from pull back of brane,  now let us derive its vertex operator in an EFT as

\beqa 
V_{\alpha}^{i}(C_{p+1},\phi)&=&i(2\pi\alpha')\mu_p\frac{1}{(p+1)!} \eps^{a_0\cdots
a_{p-1} b}
  p_b C^{i }_{a_0\cdots a_{p-1}}\Tr(\lambda_{\alpha}).
\labell{Fey4} \eeqa

  where $\lambda_{\alpha}$ is an Abelian matrix. The  off-shell 's scalar field  is Abelian and  we need to consider  two permutations as $ \Tr(\lam_2\lam_3\lam_1\lambda_{\beta}),
\Tr(\lam_2\lam_3\lambda_{\beta}\lam_1 ) $ to be able to derive  $ V_{\beta}^{j}(\phi,\phi_1,T_2,T_2)$   from  \reef{hdts212} as below 

\beqa V_{\beta}^{j}(\phi,\phi_1, T_2,T_2)&=&\xi_{1}^j \frac{1}{2}(s')(t')\ (-2i
T_p\pi)(\alpha')^{n+m+3}(a_{n,m}-b_{n,m}) \Tr(\lam_1\lam_2\lam_3\lambda_{\beta}) 
\labell{verpptt}\\&&\bigg(
(k_2\inn k)^n(k_1\inn k_2)^m+(k\inn k_2)^n (k\inn
k_3)^m  +(k_3\inn k)^n(k_2\inn k)^m+(k_3\inn k)^n(k_3\inn
k_1)^m\nonumber\\&& +(k_1\inn k_3)^n(k_2\inn k_1)^m+(k_3\inn
k_1)^n(k_3\inn k)^m +(k_2\inn
k_1)^n(k\inn k_2)^m+(k_2\inn k_1)^n(k_3\inn k_1)^m\bigg)\nonumber \eeqa

 where    $k$  is once more the momentum of 
 off-shell scalar field and  $b_{n,m}$'s are 
symmetric \cite{Hatefi:2010ik}, lets list some of the coefficients as  appeared \footnote{
 We have used 
  $k_3\inn k=k_2.k_1-k^2, k_2\inn k=k_1.k_3-k^2 $ , also overlooked all the contact terms that $k^2$ caused.
}  in \cite{Hatefi:2012wj}
\beqa
&&a_{0,0}=-\frac{\pi^2}{6},\,b_{0,0}=-\frac{\pi^2}{12},a_{1,0}=2\z(3),\,a_{0,1}=0,\,b_{0,1}=-\z(3),a_{1,1}=a_{0,2}=-7\pi^4/90,\nonumber\\
&&a_{2,2}=(-83\pi^6-7560\z(3)^2)/945,b_{2,2}=-(23\pi^6-15120\z(3)^2)/1890,a_{1,3}=-62\pi^6/945,\nonumber\\
&&\,a_{2,0}=-4\pi^4/90,\,b_{1,1}=-\pi^4/180,\,b_{0,2}=-\pi^4/45,a_{0,4}=-31\pi^6/945,a_{4,0}=-16\pi^6/945,\nonumber\\
&&a_{1,2}=a_{2,1}=8\z(5)+4\pi^2\z(3)/3,\,a_{0,3}=0,\,a_{3,0}=8\z(5),b_{1,3}=-(12\pi^6-7560\z(3)^2)/1890,\nonumber\\
&&a_{3,1}=(-52\pi^6-7560\z(3)^2)/945, b_{0,3}=-4\z(5),\,b_{1,2}=-8\z(5)+2\pi^2\z(3)/3,\nonumber\\
&&b_{0,4}=-16\pi^6/1890.\eeqa

 Now if we replace all the above vertices in the the above field theory  sub amplitude we get to produce all order singularities in effective field theory as follows
\beqa
&&2i\mu_p\frac{\eps^{a_{0}\cdots a_{p-1}b}\xi_{i} p_b C^{i}_{a_0\cdots
a_{p-1}}}{(p+1)!(s'+t'+u)}\Tr(\lam_1\lam_2\lam_3)
\sum_{n,m=0}^{\infty}(a_{n,m}-b_{n,m})[s'^{m}t'^{n}+s'^{n}t'^{m}]
  s't' \label{amphigh82}\eeqa

Now let us compare all order  EFT singularities with string theory poles that appeared in \reef{mm222}.
First of all we try to cancel off all the common factors that appeared in both EFT and string theory side.

The next step is to set $n=m=0$ so that from  the string amplitude we get to find out the following coefficient  
 $ l_{0,0} s't' =-s' t' \frac{\pi^2}{3}$ where as for the zeroth order of $\alpha'$ or leading singularity in the field theory amplitude  we obtain
 
 \beqa
4s't'(a_{0,0}-b_{0,0})&=&-s' t' \frac{\pi^2}{3}\nonumber\eeqa
 so  as we can see both coefficient match at zeroth order . Let us keep counting . 
At first order of   $\alpha'$  string amplitude proposes to us the the following coefficient  
 $ l_{1,0} s't'(s'+t')= 8\z(3)(s'+t') s't'$ 
 
while for the  $\alpha'$ or next to the leading singularity in the field theory amplitude  we gain the following term
\beqa
2 s't'(s'+t')(a_{1,0}+a_{0,1}-b_{0,1}-b_{0,1})&=&8\z(3)(s'+t') s't'\nonumber\eeqa
Let us compare at  $(\alpha')^2$ order, so that the  string amplitude  imposes to us  the following coefficient  

\beqa
l_{2,0}s't'(s'+t')^2+l_{0,1}(s't')^2  \nonumber\eeqa
meanwhile the field theory amplitude predicts the following counterparts 
\beqa
&&4 (s't')^2(a_{1,1}-b_{1,1}) +2s't'(a_{0,2}+a_{2,0}-b_{0,2}-b_{2,0})[(s')^2+(t')^2]\nonumber\\
&&=s't'(-\frac{7\pi^4}{45}(s'+t')^2+\frac{\pi^4}{45}s't')=(l_{2,0} (s'+t')^2+l_{0,1}s't')s't'\nonumber\eeqa
One can check to see  the other orders are also matched  and this clearly confirms that we have been able to precisely produce all infinite $(u+s'+t')$ bulk singularities of string amplitude in an effective field theory side where the important points were 
to look for new coupling that has been shown up in \reef{esi897} and also the fact that all order  $\alpha'$ higher derivative corrections of two scalars and two tachyons of brane anti brane are exact and correct and do differ from the corrections of non-BPS branes.
Notice that
 all  terms like $D\phi^{(1)i}. D\phi^{(2)}_i$'s  have to appear inside the all order $\alpha'$ corrections, for instance , look at the   $b_{n,m}$  coefficients in \reef{hdts212}.  Ultimately all order contact interactions could be explored from the last section of \cite{Hatefi:2012cp}.

\vskip.1in

Note that the nature of the $(s'+t'+u)$ poles can be understood as follows. It is a closed string that is absorbed by the brane and becomes an excited open string that later on decays into infinite $n$  massless scalar strings. The case $n=0$ is pure absorption, $n=1$ is mixing and $n=2$ is the origin of Hawking radiation and we have studied all the other cases. Similar poles appear also in scattering of closed strings off D-branes \cite{D'Appollonio:2015xma}.  One might be also interested in  exploring several consistency conditions for branes, various  anomalies, tadpoles, and topological couplings where we recommend some relevant papers in \cite{Blum:1997fw} and references therein.


\vskip.1in

Now we  do want to deal with all the poles in   ${\cal A}_{2}$,  note that in this case $\eps^{a_{0}\cdots a_{p}}  p^i C_{a_{0}\cdots a_{p}} = \eps^{a_{0}\cdots a_{p}}  H^{i}_{a_{0}\cdots a_{p}} $
and in order to produce the first bulk singularities , one needs to consider the following Feynman rule as 

\beqa
{\cal A}&=&V_i(C_{p+1},\phi^{(1)})G_{ij}(\phi)V_j(\phi^{(1)},T_1,T_1,\phi^{(1)})\nonumber\\&&
+V_i(C_{p+1},\phi^{(2)})G_{ij}(\phi)V_j(\phi^{(2)},T_1,T_1,\phi^{(1)})\labell{amp44}\eeqa

 also the Taylor expansion of scalar field through Chern- Simons coupling is needed   as below

 \beqa
\frac{1}{(p+1)!} (2\pi\alpha') \int_{\Sigma_{p+1}} \partial_i C_{a_0... a_{p}}  (\phi^i _{1}-\phi^i _{2})\epsilon ^{a_0... a_{p}}
 \nonumber \eeqa
 
 where  $\phi$ in the propagator  can be  $\phi^{(1)}$ and $\phi^{(2)}$ and all the effective field theory vertices are  

\beqa
G_{ij}(\phi) &=&\frac{i\delta_{ij} \delta_{\alpha\beta}}{(2\pi\alpha')^2 T_p
\left(u+t'+s'\right)}\nonumber\\
V_i(C_{p+1},\phi^{(1)})&=&i\mu_p(2\pi\alpha')\frac{1}{(p+1)!}\epsilon^{a_0\cdots a_{p}} p^i C_{a_0\cdots a_{p}}\Tr(\lambda_\alpha)\nonumber\\
V_i(C_{p+1},\phi^{(2)})&=&-i\mu_p(2\pi\alpha')\frac{1}{(p+1)!}\epsilon^{a_0\cdots a_{p}}p^i C_{a_0\cdots a_{p}}\Tr(\lambda_\alpha)\nonumber\\
V_j(\phi^{(1)},T_1,T_1,\phi^{(1)})&=&-2iT_p(2\pi\alpha')\xi_j \Tr(\lambda_1\lambda_2\lambda_3\lambda_\beta)\nonumber\\
V_j(\phi^{(2)},T_1,T_1,\phi^{(1)})&=&2iT_p(2\pi\alpha')\Tr(\lambda_1\lambda_2\lambda_3\lambda_\beta)\xi_j\labell{Fey}
\eeqa

If we apply all the  vertices into  \reef{amp44}  which is  the effective field theory rule we obtain 

\beqa
{\cal A}&=&\frac{4i\mu_p}{(p+1)!(u+s'+t')}\eps^{a_{0}\cdots a_{p}} p^i C_{a_{0}\cdots a_{p}}\xi^i\labell{amp6}\eeqa

thus we have produced the first $(u+s'+t')$ singularity pole  of ${\cal A}_{2}$.  Note that, this section does  clarify the fact that the  following  interaction coupling term 

\beqa
D\phi^{(1)i}.D\phi^{(2)}_i\label{mmbk}\eeqa

has to be appeared in an EFT. Indeed if we do not take into account   $D\phi^{(1)i}.D\phi^{(2)}_i$ term in an effective field theory 
 then $V_j(\phi^{(2)},T_1,T_1,\phi^{(1)})$  will not have any contribution to field theory and hence we are no longer able to even start producing the first  simple scalar pole of string amplitude in the effective field theory.

 The important point that is worth highlighting is that  the new coupling  \reef{mmbk} is not embedded  inside the  ordinary trace effective action. Eventually  in sections 3.2 and 3.3 of \cite{Hatefi:2012cp}
 we have shown how to get to all infinite scalar poles as well as all the contact interactions.

Finally  it would be  remarkable to find out the S-matrix of one closed string RR and four tachyons in the world volume of brane anti brane system, where by doing so, one might be able to make various remarks on the  part of the symmetrized trace tachyonic DBI effective action \cite{Hatefi:2016eh} as well as explore some kinds of different singularity structures. We hope to come over those open questions in near future.

\section*{Acknowledgments}

The author would like to thank L. Alvarez-Gaume, P. Anastasopoulos, I. Antoniadis,   N. Arkani-Hamed, C. Bachas, I. Bena, Massimo Bianchi, A. Brandhuber, G. Dvali, C. Hull ,N. Lambert, W. Lerche, K.S. Narain, J. Polchinski,  A. Rebhan, R. Russo, A. Sagnotti, J. Schwarz, A. Sen, W. Siegel, S. Thomas, A. Tseytlin, P. Vanhove, G. Veneziano and E. Witten for several useful discussions and comments. 
Some parts of this work have been performed at CERN, theory division in Switzerland as well as in ICTP,  IHES in France and  this work was partially  supported by the FWF project P26731-N27. The author would also like to thank them for providing hospitality as well as the very best scientific  environments that caused finishing this work.


\end{document}